# Photochemical Hydrogen Storage with Hexaazatrinaphthylene (HATN)

Olaf Morawski*[a], Paweł Gawryś[a], Jarosław Sadło[b] and Andrzej L. Sobolewski[a]

[a] Institute of Physics, Polish Academy of Sciences, Al. Lotników 32/46, 02-668 Warsaw, Poland

[b] Institute of Nuclear Chemistry and Technology, ul. Dorodna 16, 03-195 Warsaw, Poland

*E-mail: morawo@ifpan.edu.pl

**Abstract**

When irradiated with violet light, hexaazatrinaphthylene (HATN) extracts a hydrogen atom from an alcohol forming a long-living hydrogenated species. The apparent kinetic isotope effect for fluorescence decay time in deuterated methanol (1.56) indicates that the lowest singlet excited state of the molecule is a precursor for intermolecular hydrogen transfer. The photochemical hydrogenation occurs in several alcohols (methanol, ethanol, isopropanol) but not in water. Hydrogenated HATN can be detected optically by an absorption band at 1.78 eV as well as with EPR and NMR techniques. Mass spectroscopy of photoproducts reveal di-hydrogenated HATN structures along with methoxylated and methylated HATN molecules which are generated through the reaction with methoxy radicals (remnants from alcohol splitting). Experimental findings are consistent with the theoretical results which predicted that for the excited state of the HATN-solvent molecular complex, there exists a barrierless hydrogen transfer from methanol but a small barrier for the similar oxidation of water.

# 1 Introduction

Hexaazatrinaphthylene (HATN, Scheme 1) is an electron poor system with high electron affinity.[1] Its derivatives are being intensively investigated as desirable alternatives for sustainable lithium-ion battery electrodes that offer high capacity and long-term cyclic stability.[2-5] HATN, when connected to an electron donor system offers applications like deep-red thermally activated delayed fluorescence (TADF) emitters.[6] Its polymeric derivatives are also considered as materials for advanced energy conversion and storage,[7] including supercapacitors,[8] as well as metal-free catalysts for electrochemical reactions like oxygen reduction, oxygen evolution or overall water splitting.[9] Containing several pyrridic-N atoms, HATN may be considered as an "improved acridine" or a material similar to graphitic carbon nitride (gCN). These organic polymer materials have recently received extensive attention as photo-catalysts for solar water oxidation.[10]

In the prevailing interpretation of photocatalytic water oxidation, charge separation in the photocatalyst is the first step of the process, whereas the reduction of protons by separated electrons and the oxidation of water by holes at solid-liquid interfaces are the last steps which complete the process. In most of the numerous studies of molecular hydrogen generation with carbon nitride photocatalysts, amines or alcohols were used as sacrificial electron donors.[11-12] In just a few cases, $H_2$ evolution without the addition of a sacrificial electron donor has been reported.[13-16] This poses a question about the role of alcohols and amines in the water oxidation process.

In an alternative scenario proposed by Domcke, Sobolewski and collaborators, the water-oxidation reaction with gCN models is related to the molecular properties of the heptazine or triazine building blocks. Theoretical studies of the photochemical reaction of hydrogen-bonded complexes of pyridine[17], acridine[18], triazine[19] and heptazine[20] with water indicate that the N-atoms of these aromatic heterocycles are the locally active sites for water oxidation. In that framework, the primary photochemical reaction is the electron transfer from a hydrogen-bonded water molecule to the photoexcited chromophore, followed by the transfer of a proton from the oxidized water to the chromophore. Experimental studies of hydrogen-bonded complexes of heptazine derivatives with hydroxylic substrate molecules (water and phenol) confirm theoretical predictions that functionalized derivatives of heptazine can photooxidize water and phenol in a homolytic photochemical reaction.[21] For pyridine, evidence for the water splitting photocatalytic reaction Py–$H_2O$ + hv → PyH• + OH• was given by the UV excitation of molecular Py–($H_2O$)$_n$ clusters obtained in a supersonic expansion and monitoring the PyH• reaction product.[22] The results unambiguously showed that PyH• is produced, and thus that water is split using pyridine as a photo-catalyst. Partial water photolysis observed for oxotitanium tetraphenylporphyrin[23] and titanyl phthalocyanine[24] indicates the photochemical homolytic reaction can also occur with organometallic catalysts.

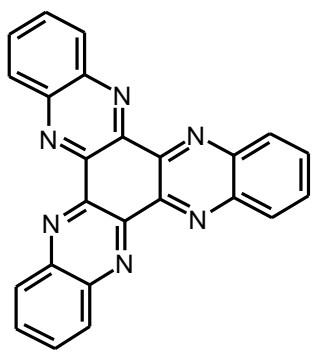

Scheme 1. **Structure of HATN.**

The HATN molecule possesses six nitrogen atoms that may form hydrogen bonds with protic solvents and is therefore a good candidate to test alternative mechanisms of alcohol or water oxidation. A recent study has shown that in solvents HATN is weakly emissive, with a fluorescence quantum yield $\phi_{Fl}$ below one percent.[25] Excited state relaxation is dominated by internal conversion of $\phi_{ic}$ ~ 0.6, and intersystem crossing, with triplet formation yield $\phi_{isc}$ ~ 0.4. In methanol the fluorescence spectrum is slightly blue-shifted indicating formation of hydrogen bonds. Fluorescence decay times of 50-77 ps in non-protic solvents, are shortened to below 20 ps in methanol where the $\phi_{Fl}$ also decreases and $\phi_{isc}$ drops down to 0.2, suggesting opening of yet another non-radiative de-excitation pathway. An excited state proton transfer process in solute-methanol complexes resulting in HATN hydrogenation may account for the photochemical reaction observed in this solvent.[25]

The present work aims to explore, theoretically as well as experimentally, the photophysics of HATN in alcohols to examine the possibility of partial splitting of methanol in the excited state reaction:

HATN + CH$_3$OH + hv → HATN-H• + CH$_3$O•     (1)

An attempt is made to check if similar excited state process occurs for water.

## 2 Results and Discussion

### 2.1 Experimental results

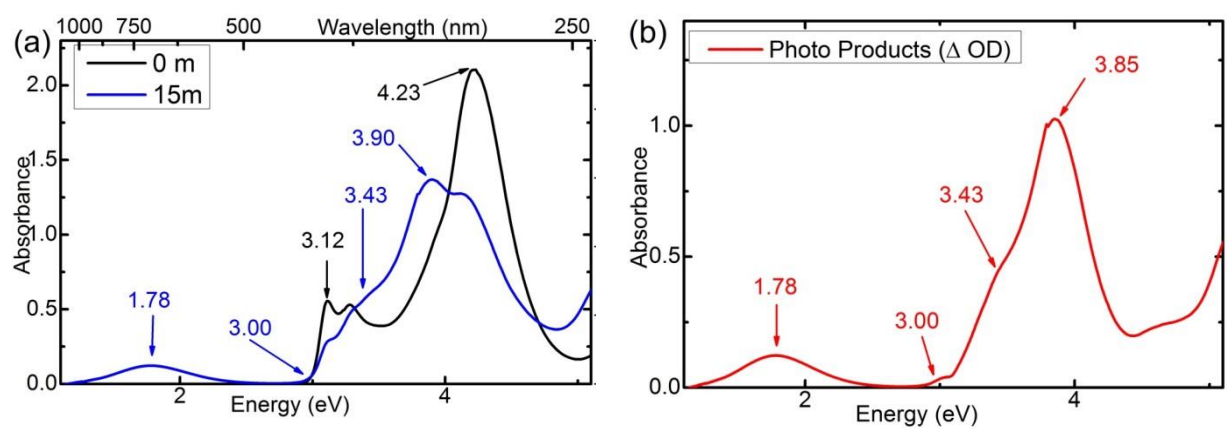

*Figure 1.* (a) Absorption of HATN solution in methanol before (black line) and after 15 minutes of phototoirradiation with a 405 nm laser (blue line), (b) absorption spectrum of photoproducts – the difference between "after-" and "before irradiation" absorption spectra of panel (a). Numbers denote band maxima in electronvolts.

Absorption and fluorescence spectra of HATN in alcohols are presented in Figure SF1 of the Supporting Information (SI). Absorption spectra in methanol, perdeuterated methanol, ethanol and 2-isopropanol are identical to that shown in Figure 1a before irradiation. Fluorescence spectra in these solvents are also similar. However, due to the low solubility of HATN, a wing on the low energy side of the emission spectra appears, especially in 2-isopropanol, indicating the presence of aggregates.[25] Photoirradiation of HATN in degassed methanol results in blue colorization of the solution, see Figure SF2, and a change in the absorption spectrum – Figure 1a. The blue color of the photoirradiated solution originates from build up of the absorption band with a maximum at 1.78 eV (696.5 nm). Figure 1b presents the absorption difference between the fresh and the 15 minutes photoiradiated solutions. From this, it is evident that the new photoproduct's absorption bands are also located in the UV region. The vibronic structure of the HATN first absorption band at 3.12 eV and the main absorption band at 4.23 eV are barely visible in the absorption spectrum after 15 min optical excitation with a 405 nm laser indicating effective conversion of HATN molecules into photoproducts. Photoexcitation of HATN in CD$_3$OD, ethanol and 2-isopropanol leads to build up of identical absorption bands at 1.78 eV (Figure SF3). In contrast, photoirradiation of a HATN suspension in water does not change the absorption spectrum (Figure SF4a) nor the color of the suspension. Same is true for a HATN suspension in water – methanol mixture (Figure SF4b).

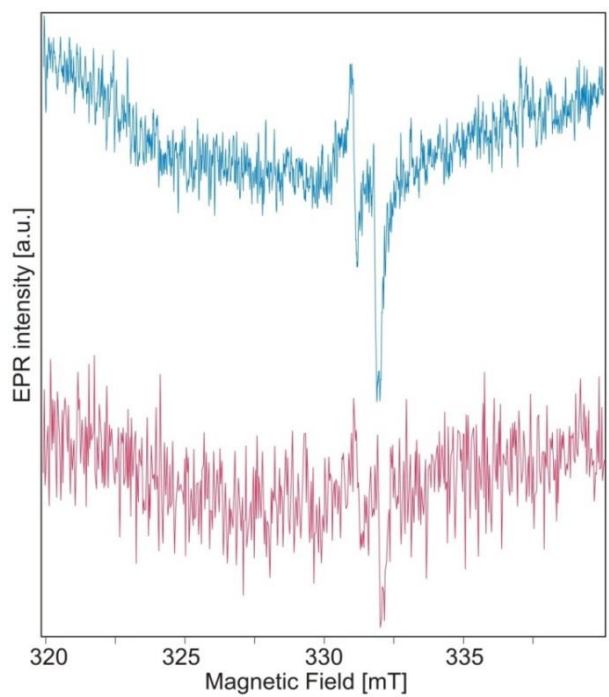

*Figure 2.* EPR spectra of HATN in methanol before (red) and after (blue line) photoirradiation with 405 nm laser. Spectra recorded at 100 K, photoirradiation conducted at room temperature.

The EPR spectrum of the photoirradiated solution (Figure 2, top) contains the signal of the ½ spin with hyperfine splitting A = 0.75 mT and g-factor g=2.0044, proving that photoexcitation of HATN in methanol leads to the creation of radicals. The EPR spectrum of the colored solution is similar to the EPR spectra observed for lithiated HATN obtained in discharging reactions.[2] Hence, the photo induced conversion of HATN can be related to redox chemistry of this azaaromatic molecule. In the EPR spectrum recorded before optical excitation, a very weak signal of radicals can be detected as well (Figure 2, bottom). It points to the presence of trace amounts of unpaired spins which might have been produced by daylight photons during the lengthy freeze-pump-thaw de-oxidation procedure or during transportation of the probe to the EPR apparatus.

NMR spectra of HATN in $CD_3OD$ change as a result of photoexcitation (Figure SF5). The non-irradiated solution contains structured bands at 8.64 and 8.16 ppm which are known as signals of HATN hydrogen atoms.[2] In spectra of irradiated solutions, several new bands build up at the cost of the HATN hydrogen features for which intensity decreases with increasing photoexcitation time. This is clearly visible in Figure SF6 where the integral intensity of all NMR bands are depicted as a function of the irradiation time. As the HATN hydrogens bands' intensity decreases, the intensity of all photoproducts bands increases upon photoexcitation, whereas intensity of the singlet at 8.53 ppm of acidic protons remains constant.

Kinetic isotope effect (KIE) is observed for fluorescence decay time, $\tau_F$, of HATN in methanol. In $CD_3OD$, $\tau_F(D) = 28$ ps whereas in $CH_3OH$ $\tau_F(H) = 18$ ps (Figure SF7), which yields the apparent KIE = 1.56, a value unambiguously pointing to the role of hydrogen transfer in the excited state relaxation pathways. The fluorescence decay rate, $k_F = 1/\tau_F$, is a sum over rates of all excited state relaxation pathways: radiative, $k_r$, internal conversion, $k_{ic}$, intersystem crossing, $k_{isc}$, and Proton Coupled Electron Transfer, $k_{PCET}$.

$$k_F = k_r + k_{ic} + k_{isc} + k_{PCET} \qquad (2)$$

Taking decay time in toluene $\tau_F = 77$ ps as a reference where PCET process does not operate,[25] and assuming that $k_r$, $k_{ic}$ and $k_{isc}$ are the same for HATN in toluene and methanol, one may calculate the rate for methanol $k_{PCET}(H) = 4.3 \times 10^{10}$ sec$^{-1}$ and deuterated methanol $k_{PCET}(D) = 2.3 \times 10^{10}$ sec$^{-1}$. This yields the true KIE = 1.87. All values are typical for the PCET process.

The excitation of the colored solution at 659 nm does not result in any emission in the 750 - 1400 nm range of wavelengths, so the relaxation of this product must be dominated by non-radiative processes. In contrast, the UV excitation of the blue solution changes the fluorescence spectrum. In parallel to the HATN spectrum with maxima at 419 and 430 nm, a new emission builds up at longer wavelengths upon photoexcitation (Figure 3a) proving some of photoproducts are fluorescent.

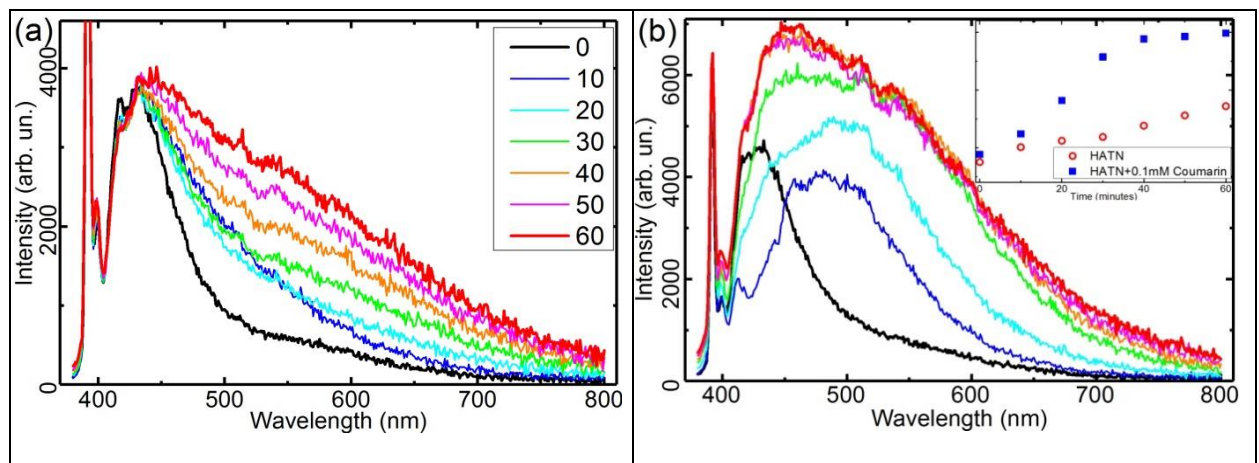

***Figure 3.*** *Fluorescence spectra of HATN in a methanol solution irradiated with a 405 nm laser and recorded with 10 minutes intervals (a), and the same for HATN with 0.1 mM coumarin in methanol (b). Spectra recorded with 340 nm excitation. Legend in (a) specifies time of irradiation (in minutes) and colors of spectra, inset in (b) depictures integral intensity of fluorescence spectra of HATN (red circles) and HATN with 0.1 mM coumarin solution (blue squares) versus time of photo-irradiation. The sharp feature at 392 nm is a Raman line of methanol.*

The appearance of new emission bands can be related to the reaction of HATN molecules with radicals generated in solution. The excited state proton transfer occurring in the HATN-methanol hydrogen bonded complex results in formation of hydrogenated HATN (HATN-H•) and methoxy (MeO•) radicals. The latter, if not scavenged, may recombine with a hydrogen attached to a nitrogen atom of HATN, attack the hydrogenated compound at another position or react with another bare HATN molecule. Thus one may expect methoxylated HATN photoproducts in the irradiated solution. To prove the presence of MeO• radicals in the solution we used coumarin, a known radical scavenger used in color dosimetry of hydroxyl radicals.[37] Bare coumarin absorbs at short wavelengths, 310 nm, and is not fluorescent because in the excited state the lactone ring opens due to the low electron density on it.[37] 7-hydroxycoumarin absorbs light at longer wavelengths, 340 nm, and becomes emissive with the maximum of the fluorescence spectrum appearing at 425 nm.[38,39] The methoxy group is a weak electron donor so one may expect that methoxylation of coumarin increases electron density on its rings which should result the molecule becoming fluorescent. Thus coumarin can serve as a fluorescent methoxyl radical scavenger as well. Results presented in Figure 3b confirms this expectation. At even 10 minutes irradiation at 405 nm the fluorescence is changed significantly and the new emission band, red shifted by 70 nm to the HATN monomer emission, builds up on the low energy tail of the fluorescence spectrum. Further 405 nm irradiation leads to a significant growth of the new photoproduct population as revealed by more than a tri-fold increase in total intensity of the emission. After 30 minutes of the 405 nm irradiation, the increase of the fluorescence intensity slows down and intensity stabilizes, pointing to a stationary equilibrium between photochemical processes.

The photoproducts in anaerobic solutions are persistent, e.g. the irradiated solution in the sealed EPR tubes retains a blue colour for several months. We therefore attempted high resolution mass spectroscopy under oxygen free conditions to determine the masses and possible structures of photochemically created compounds. Measurements were performed before and after irradiation with positive and negative polarity. In the fresh solution HATN monomers, dimers and trimers were found at the respective mass to charge ratios (m/z): 407.1 [HATN+Na]$^+$, 791.21 [2HATN+Na]$^+$ and 1157.32 [3HATN+Na]$^+$ (Figures SF8 – SF11), an unsurprising result at the 0.1 mM concentration used, and consistent with results found through optical spectroscopy means earlier.[25] In the irradiated blue solution, several new structures have been isolated, four of the most abundant are presented in Scheme 2 and figures SF12-SF15, the other photoproducts and alternative structures are depicted in figures SF16-SF17. Observation of di-hydrogenated HATN is the main finding, confirming the excited state intermolecular hydrogen transfer reaction (1). The presence of methoxylated HATN is another confirmation of reaction (1) and validates the indirect results of the color dosimetry method discussed above. It proves that bare coumarin can be used as a methoxyl radical scavenger and a fluorescence probe in the color dosimetry of these radicals. It also indicates the methanol split reaction (1) could not be fully reversible, as a fraction of the photocatalyst molecules are converted to new structures in a reaction with radicals. Observation of the methoxylated di-hydrogenated HATN reveals the photocatalyst can retain hydrogen atoms even when attacked by radicals. Other photoproducts provide evidence for several 'dark chemistry' reactions in the solution triggered by methoxy radicals. Di-deuterated HATN obtained in deuterated methanol (Figures SF18, SF19) points to the solvent as the source of hydrogens/deuteriums in the hydrogenation/deuteration process. Observation of HATN-OCD$_3$ in deuterated methanol (Figure SF20) complements the proof for the reaction (1).

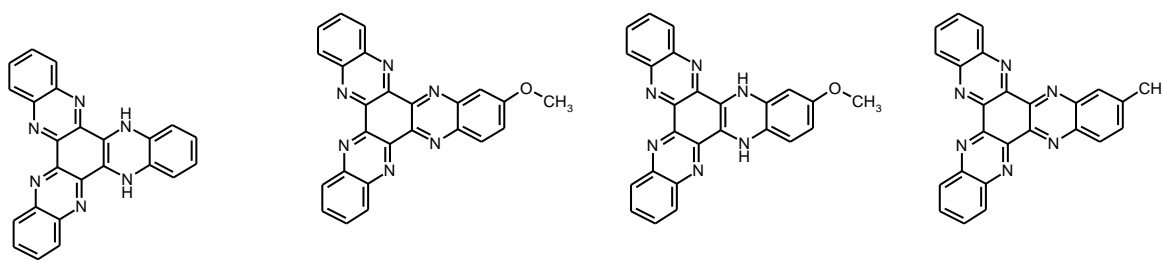

**Scheme 2.** *Structures of the main HATN photoproducts found in the photoirradiated methanol solution with high resolution mass spectrometer. Positions of the attached atoms and groups are ambiguous.*

Similarly to hydrogen storage in monomers, HATN dimers and trimers are also hydrogenated during the photoreaction and abstract four and six H atoms respectively as a comparison of figures SF8 and SF12 reveals. In Figure SF12 along with the hydrogenated monomer line at m/z=409.11 [(HATN+2H)+Na]$^+$ the lines for hydrogenated dimers are visible at m/z ~795 and a group of small intensity lines in the m/z ~ 1177 - 1181 range can be assigned to trimers. Detailed analysis of dimers (Figure SF21), reveals the presence of two structures: a dimer with two hydrogens [2(HATN+H)+Na]$^+$ at m/z=793.22 and a dimer with four hydrogens [2(HATN+2H)+Na]$^+$ at m/z=795.24. The isotope profiles overlaid in Figure SF22 clearly confirm this finding. For trimers, three structures have been obtained: with two, four and six H atoms at m/z ratio 1177.32, 1179.35, and 1177.32 respectively (Figures SF23 and SF24). This reveals the hydrogenation occurs in a stepwise process with close shell structures being stable whereas the radical forms of odd number of hydrogen atoms subject very likely to attachment of another H-bonded atom. This may explain the very weak EPR signal presented in Figure 2 and raises a question about the elementary steps of the photochemical reaction - whether in aggregates the hydrogenation occurs in concerted two-H atoms transfer or in a very quick single PCET by single PCET successive steps. The results obtained for small aggregates suggests that the process may also occur in larger aggregates and potentially even at the molecular surface, which could open a way for practical applications for the photochemical hydrogenation mechanism.

It is known that electrochemical oxidation of polyazaacenes in an aprotic electrolyte is accompanied by proton loss whereas reduction(s) is(are) followed by a self-protonation reaction and consists of separate steps.[40] A derivative of tetraazapentacene undergoes two successive reversible reductions and its electrochemistry shows facile reducibility.[41] For HATN, successive six-fold lithiation has been reported in the solid state,[2] setting expectations for the electrochemical reactions in alcohols. Our cyclic voltammetry (CV) measurements for HATN in methanol aimed to check how many reductions are possible in this protic solvent. The result in methanol is presented in Figure 4. The two close reduction peaks are observed at -0.84 and -0.92 V, the oxidation peaks are positioned at 0.87 and 1.24 V. The electrochemical hydrogenation of HATN is reversible and may subsequently occur on more than one redox site. The difference between reduction and oxidation potentials is 1.73 eV close to the 1.78 eV maximum of the red absorption band of the photoirradiated HATN solution. Comparable HOMO - LUMO energy gap of the blue photoproduct and electrochemically hydrogenated HATN provides another argument for the photochemical hydrogenation of HATN in alcohols.[42] Interestingly, the reaction path ends up with di-hydrogenated HATN. This indicates the attachment of the third hydrogen must be energetically unfavoured.

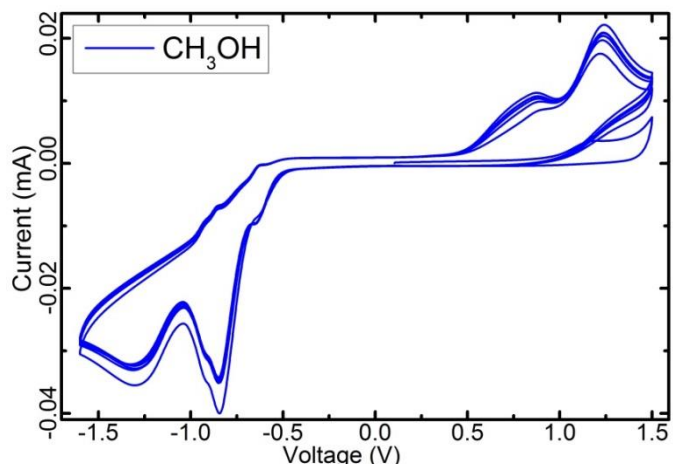

**Figure 4.** *Cyclic voltammograms of HATN in methanol recorded versus Ag/Ag$^+$ electrode with 100 mV/s scan rate.*

In contrast to several experimental confirmations for the hydrogenation of HATN in methanol there is no evidence for the process in water. Its absence may result from several factors: very low HATN solubility in water, its hydrophobicity or high dielectric constant of water. HATN was proved to be hydrophobic[43], a feature that affects a solvation of a solute and also the PCET process.[44] High dielectric constant of water changes the Born energy $E = e^2/8\pi\varepsilon_0\varepsilon a$ what influences the energetics of PCET reaction. Calculation may bring insight into the actual mechanism of the hydrogenation.

### 2.2 Theoretical exploration

An isolated HATN molecule exhibits $D_{3h}$ symmetry in the electronic ground state. The vertical energy of several lowest excited states adapted from Ref. 25 and computed with the aid of the ADC(2)/cc-pVDZ method are presented in Table 1.

**Table 1**. *Vertical excitation energy (in eV) and oscillator strength (in parentheses) of the lowest excited states of HATN and its complexes with water and methanol molecules determined with the ADC(2)/cc-pVDZ method at the MP2/cc-pVDZ equilibrium geometry of the ground state.*

| State | HATN[a] | HATN-H$_2$O | HATN-MeOH |
|---|---|---|---|
| $^3A(\pi\pi^*)$ | 2.80 | 2.78 | 2.77 |
| $^3E(\pi\pi^*)$ | 2.98 | 2.95 | 2.94 |
|  |  | 2.96 | 2.95 |
| $^3E(n\pi^*)$ | 2.99 | 3.01 | 3.01 |
|  |  | 3.02 | 3.01 |
| $^3A(n\pi^*)$ | 3.08 | 3.12 | 3.11 |
| $^1E(n\pi^*)$ | 3.41(0.0) | 3.43(0.0) | 3.42(0.0) |
|  |  | 3.43(0.0) | 3.43(0.0) |
| $^1A(n\pi^*)$ | 3.50(0.0) | 3.53(0.005) | 3.52(0.005) |
| $^1A(\pi\pi^*)$ | 3.70(0.0) | 3.70(0.072) | 3.69(0.066) |
| $^1E(\pi\pi^*)$ | 3.71(0.422) | 3.67(0.103) | 3.66(0.101) |
|  |  | 3.68(0.217) | 3.67(0.217) |

[a] Adopted from Ref. 25.

The nondegenerate $S_1$(A) and the degenerate $S_2$(E) excited states are of $n\pi^*$ electronic nature and are followed by several higher energy singlet states of $\pi\pi^*$ origin. The first three excited singlet states are "dark" for absorption from the ground state, and the lowest absorbing singlet state is located at 3.71 eV. The computed absorption spectrum of HATN (Figure SF25b) is dominated by an absorption band to the fully allowed higher singlet state located at 4.58 eV at this level of theory. Assuming the shift of 0.4 eV, a typical value for energy overestimation at this level of theory, the computed absorption spectrum is in a fair agreement with experiment (Figure 1). Below the respective singlet states, the triplets are located with a similar pattern of nondegenerate and degenerate $\pi\pi^*$ and $n\pi^*$ excited states. However in the triplet manifold, the $n\pi^*$ states are above the $\pi\pi^*$ states because the exchange integral is larger for the latter ones.

The six peripheral nitrogen atoms of HATN can serve as acceptor atoms for hydrogen bonding with a protic solvent. Complexation of HATN with a protic molecule (methanol or water) breaks the $D_{3h}$ symmetry and results in a small hypsochromic shift of $n\pi^*$ and a bathochromic shift of $\pi\pi^*$ states

(Table 1). Complexation with three such solvent molecules, represents formation of the first solvation shell, and straightens this effect (see Table ST1 and Figure SF25b).

ADC(2)/cc-pVDZ geometry optimization of the HATN-CH$_3$OH complex in the lowest excited (singlet and triplet) states results in determination of a minima in the states of the ππ* electronic nature, but leads to barrierless oxidation of a methanol molecule in the states of nπ* electronic origin. In order to investigate these processes in more detail, we define the bond length $R_{OH}$ of the OH group of the methanol molecule involved in the hydrogen bonding as the reaction coordinate for the H-atom transfer reaction from the methanol molecule to HATN and determine the minimum potential-energy (PE) functions of the lowest ππ* and nπ* excited states by geometry optimization of the complex for fixed value of $R_{OH}$. The resulting PE profiles for the HATN-CH$_3$OH complex are shown in Figure 5.

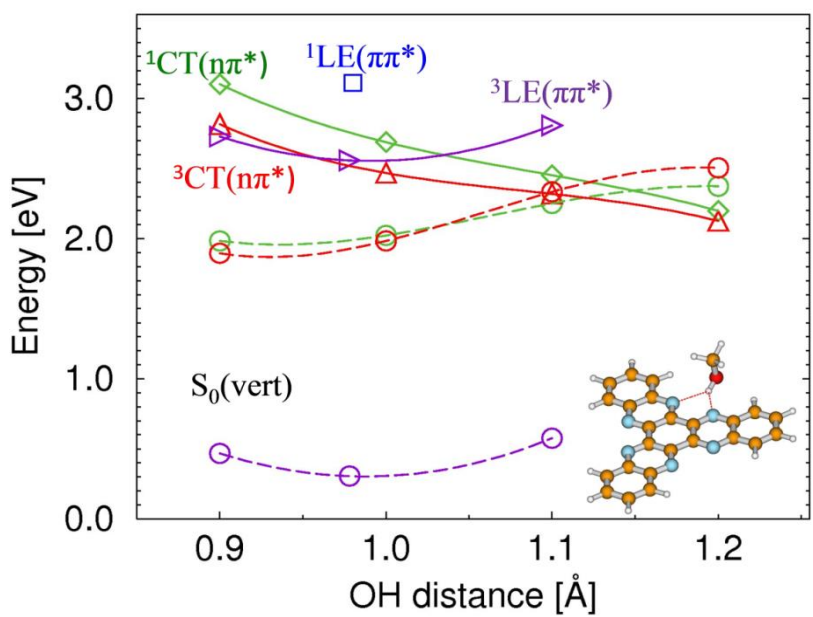

*Figure 5.* PE profiles of relaxed scan for H-atom transfer from methanol to HATN computed along the OH distance of MeOH in different electronically excited states of the complex: $^1$ππ*- blue squares, $^1$nπ*- green diamonds, $^3$ππ*- violet triangles, $^3$nπ*- red triangles, connected by solid lines. Vertical energy profile of the ground state (circles connected by dashed lines) was computed along the relaxed scan in a given electronic state as encoded by color.

It can be seen upon inspection of Figure 5 that the adiabatic minimum of the $^1$ππ* state lies slightly above the $^1$nπ* state and an incremental stretching of the OH bond in the former state results in a non-adiabatic collapse of the $^1$ππ* state to the nearby lying $^1$nπ* state that results in a barrierless H-atom transfer to the nitrogen atom of HATN. The PE profile of the $^3$nπ* state varies along the $R_{OH}$ coordinate in a parallel fashion to the respective singlet state and also results in barrierless dissociation of the methanol molecule.

When the energy of the nπ* states is minimized by geometry optimization, these states not only become the lowest excited state of the complex beyond ROH =1.1 Å, but for further stretching of the OH bond length their energy even drops down below the vertical energy of the closed-shell ground state. The crossings of the PE of the excited $^{1,3}$nπ* states (solid green and red lines in Figure 5) with the PE of the closed-shell ground state (dashed lines) represent conical intersections of electronically excited states with the ground state. Such crossings are not observed for the excited state of the $^3$ππ* electronic nature where vertical energy of the ground state is separated by large energy gap.

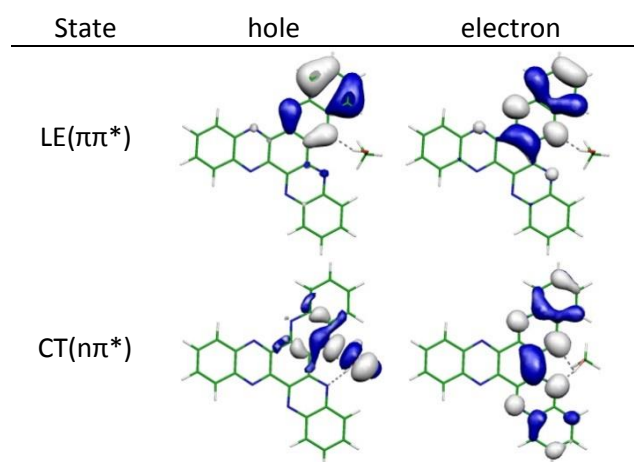

| State | hole | electron |
|---|---|---|
| LE(ππ*) | | |
| CT(nπ*) | | |

*Figure 6. Natural Transition Orbitals determined at the minimum the LE(ππ*) and at OH=1.0Å of the CT(nπ*) states of the HATN-CH₃OH complex with the aid of the ADC(2) method.*

Inspection of Natural Transition Orbitals (NTOs, Figure 6) determined at the minima of the lowest ππ* states and at $R_{OH}$ =1.0 Å of the lowest nπ* states (Figure 5) shows that the ππ* states can be classified as the locally-excited (LE) states of HATN, while the nπ* states are of the charge-transfer (CT) character where an electron is transferred from a methanol molecule to HATN. Since protons follow electrons on a barrierless PE surface of the nπ* states the HATN-CH₃OH complex represents a classical example of the proton-coupled electron-transfer (PCET) process accounted recently for numerous other hydrogen-bonded systems [17-21] resulting in a barrierless formation of a new chemical species which is a hydrogen-bonded HATN-H•···CH₃O• biradical which is more stable than the close-shell ionic species (HATN-H⁺···CH₃O⁻).

A similar pattern of excited-state PE profiles were determined for the HATN-H₂O complex and is presented in Figure SF28 of the SI. Unlike in the HATN-CH₃OH complex discussed above, the minimum of the ¹ππ* state is shallow and the stabilization of the CT(nπ*) states with respect to the LE states of the ππ* orbital nature is smaller. In HATN complexed with water the lowest ³LE state represents only a meta-stable minimum intersected by PE profile of dissociative ³CT state. On the other hand, the ³ππ* state is stable against water oxidation in a much larger range of the OH distance (up to $R_{OH}$ =1.2 Å) than in the HATN-MeOH complex discussed above. NTOs determined for the lowest excited states of the HATN-H₂O complex shown in Figure SF27 are qualitatively similar to the respective NTOs of the HATN-CH₃OH complex.

In both, HATN-CH₃OH and HATN-H₂O complexes, the energy of the ππ* states decrease whereas the energy of the nπ* states increase with respect to the values of the bare HATN molecule in the gas phase (Table 1). This effect is typical for heterocyclic molecules in protic solvents. However, in complex with a single water molecule the adiabatic minimum of the ¹LE(ππ*) state is nearly isoenergetic with ¹CT(nπ*) and the minimum of the ³LE(ππ*) state is already below the ³CT(nπ*) state, whereas in complex with a single methanol molecule the minimum of the ¹LE state is well above ¹CT state and the ³LE state is quasi-degenerate with the ³CT state. Thus solvation effects in bulk water would dip the ¹ππ* state minimum below the reactive ¹nπ* state resulting in a trap for the PCET process in the ¹ππ* state, whereas in methanol the 0.4 eV energy gap between ¹ππ* - ¹nπ* states very likely preserves a barrierless ¹ππ* to ¹nπ* non-adiabatic transition and hydrogen transfer from methanol to HATN in the latter state. An apparent quasi-degeneracy between the bound ³ππ* state and the dissociative ³nπ* state, makes this reaction also possible in the triplet manifold of the complex.

PE profiles of the HATN-MeOH complex discussed above, result in formation of the radical pair HATN-H•···MeO• due to an optical excitation. The spectroscopic effect of the reaction is the colorization of the degassed solution of HATN in methanol (Figure SF2). The blue color of the solution originates from appearance of new absorption band with maximum at 1.78 eV as discussed in the experimental section (696.5 nm, Figure 1). The main candidate for the primary photoproduct, is hydrogenated HATN (HATN-H•) formed in the reaction (1) discussion preformed above. This is also consisted with the EPR spectrum of the photoirradiated solution (Figure 2). The ADC(2)/cc-pVDZ simulated absorption spectra of HATN-H• radical is shown in Figure SF26, and indeed a new absorption band below 2 eV appears.

The radical species are expected to be short lived and can be eliminated from the solution either due to geminate recombination with a methoxy radical (OCH$_3$•) or due to the disproportionation reaction:[45]

2HATN-H• → HATN + HATN-2H                    (3)

The reaction (3) is exothermic with enthalpy ΔE=-0.57 eV estimated at the D3-DFT/B3-LYP/cc-pVDZ level of theory. The ADC(2) simulated absorption spectrum of the lowest-energy close-shell product of the reaction (3) shown in Figure 7 is qualitatively similar to the spectrum of the HATN-H• radical shown in Figure SF26 since both species possess an absorption band close to 2 eV.

One can wonder if such a disproportionation reaction can be continued resulting in formation of HATN-4H or even HATN-6H species. This is rather unlikely since at the DFT level of theory such reactions are endothermic. The ADC(2) simulated absorption spectrum of HATN-4H is shown in Figure SF26 and possess a new absorption band around 2.5 eV which is apparently absent in the spectrum of the photoproduct (Figure 1b).

In order to check the methodological sensitivity of the simulated absorption spectrum of the most likely photoproduct (HATN-2H), in Figure 8 the absorption spectrum of the lowest energy isomer of HATN-2H computed with three different ab initio method are shown. It is seen that all three methods predict an absorption band below 2 eV, then a transparent window between 2 and 3 eV, followed by several absorption bands of increasing intensity. Thus the computed absorption spectrum of HATN-2H reproduces all the crucial features observed for the photoproduct of HATN in methanol solution.

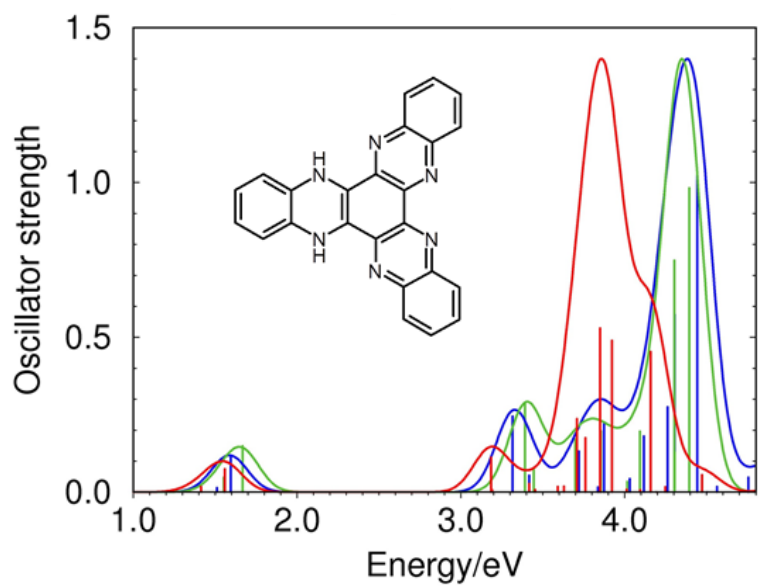

*Figure 7.* Simulated absorption spectra of HATN-2H computed with TD-DFT (red), ADC(2) (blue), CC2 (green) methods. The computed stick spectra were convoluted with Gaussian function of 0.25 eV FWHM.

We next sought to check if the confirmed product structures can be assigned to new bands observed in the $^1$H NMR spectrum of the irradiated solution. For that, DFT $^1$H magnetic shielding constants were calculated for several closed-shell compounds; doubly hydrogenated HATN, hydrogenated and methoxylated HATN (Figure SF29) as well as bare HATN for a reference. Using the linear relationship between the shielding and experimental chemical shift [46] an attempt to fit the experimentally observed $^1$H lines of HATN photoproducts with theoretically obtained values was made. The lines calculated for the HATN-2D(2) and HATN-D-CD$_3$(1) structures have close experimental counterparts (Figure SF30). However some experimentally observed bands cannot be reconstructed, so other closed-shell structures or long living radicals must contribute to the bands of products as well.

Dihydrogenated HATN (HATN-2H) formed according to reactions (1) and (3) is an energetic compound with energy stored in the reducing potential of hydrogen atoms. Potentially, the energy can be utilized (with a proper catalyst) for formation of molecular hydrogen, or can be used in a fuel cell according to the reaction:

2HATN-2H + O$_2$ → 2HATN + 2H$_2$O          (4)

The enthalpy of this reaction, estimated at the D3-DFT/B3-LYP/cc-pVDZ level of theory, is ΔE=-70.4 kcal/mol (-3.05 eV) providing -17.5 kcal/mol (-0.76 eV) per hydrogen atom. Indeed, a simple experiment confirms this result - the blue color of the photoirradiated solution quickly diminishes and the cuvette becomes transparent when oxygen may penetrate it. The control over such a process and construction of a cell that could be used for practical purposes is, certainly, beyond the scope of this work. However, to provide some insight into a process of hypothetical fuel cell making use of di-hydrogenated HATN, energetics of HATN-2H oxidation in comparison to the first step of methanol oxidation is briefly discussed in the theoretical part 2.2 of the SI.

## 3 Conclusions

Our results point to the conclusion that photochemical hydrogenation of HATN in methanol and other alcohols is possible with visible violet light. The process occurs in alcohols but not in water where hydrophobic HATN forms suspension and also the energetic condition for intermolecular excited-state PCET reaction is unfavorable. The excited state hydrogen transfer leads to stable di-hydrogenated HATN monomers which are abundant in the mass spectroscopy spectrum. The population of single hydrogenated molecules is small and results in a very weak EPR signal. Other radical forms, e.g. triple hydrogenated HATN, were not observed with any of techniques deployed in this work. The photochemical hydrogenation also occurs in small aggregates, were up to two H atoms can be retained per HATN molecule. Thus large HATN aggregates may potentially be used for high-density hydrogen storage. The persistency of the obtained energetic compounds can open a new strategy to photocatalytic hydrogen generation and storage.

## 4 Experimental Section

### 4.1 Experimental methods

The synthesis and characterization of HATN has been previously described.[25] For this work the material was purified by several crystallizations from spectroscopic grade dichloromethane. Spectroscopic grade methanol, ethanol, isopropanol and HPLC grade CD$_3$OD, all from Sigma-Aldrich were used as received for the optical, EPR, NMR and mass spectroscopy measurements.

Room temperature measurements were performed with dilute solutions in standard quartz cells (10 × 10 mm). Solutions were N$_2$ purged and stoppers were wrapped with laboratory film to slow down the oxygen penetration. Absorption spectra at room temperature (21°C) were recorded using a PerkinElmer Lambda 35 spectrophotometer. Emission spectra were obtained using a FLS 1000 of Edinburgh Instruments spectrofluorometer. For near IR emission spectra in the 750 - 1400 nm range of wavelengths a Hamamatsu C9940-02 nitrogen cooled photomultiplier was used as a detector for the FLS 1000 spectrofluorometer. Fluorescence kinetics studies were performed using the time correlated single photon counting technique. A mode-locked Coherent Mira-HP picosecond laser pumped by a Verdi 18 laser was used for excitation. The fundamental pulses of the Mira laser (tunable within 760 - 800 nm) were up-converted to ~390 nm. The temporal width of the excitation pulses was ~280 fs and the instrument response function (IRF) about 40 ps. Fluorescence was dispersed with a 0.25 m Jarrell-Ash monochromator and detected with a HMP-100-07 hybrid detector coupled to an SPC-150 PC module, (Becker&Hickl GmbH). Fluorescence decays were analyzed with deconvolution software using a nonlinear least squares procedure with the Marquardt algorithm.[26] A standard $\chi^2$ test and Durbin-Watson (DW) parameter along with residual and autocorrelation function plots were used to assess the quality of a fit. The estimated accuracy for the determination of decay time was below 10 ps.

For EPR spin measurements the samples were placed in suprasil EPR tubes. A EMXplus EPR Bruker CW X-band spectrometer equipped with an ER 4131VT nitrogen cryostat was used to measure spectra at 100 K. The following parameters were used: 1 mW microwave power, 0.1 mT modulation amplitude with 100 kHz modulation frequency, 2.560 ms time constant and 20 ms conversion time. An accurate g-value determination was done using a polycrystalline DPPH standard (g = 2.0036). The samples were degassed with freeze – pump – thaw method and sealed to block the molecular oxygen.

$^1$H FT-NMR spectra were recorded using Bruker AVANCE 500 MHz spectrometer. The chemical shifts are given with the residual solvent peak of methanol-$d_4$. The sample was a saturated solution of HATN which was further transferred to NMR tube and efficiently purged with argon.

Mass spectrometry analyses were performed using Ultra-Performance Liquid Chromatograph ACQUITY UPLC I-Class (Waters) coupled with Synapt G2-S mass spectrometer (Waters) equipped with the electrospray ion source and quadrupole-Time-of-flight mass analyzer. The resolving power of the TOF analyzer was set to 20000 FWHM. The instrument was controlled and recorded data were processed using the MassLynx V4.1 software package (Waters). Both the positive and the negative ion modes were used in mass spectrometry. The measurements in positive mode were performed with capillary voltage set to 3.00 kV. The desolvation gas flow was 700 L/h and temperature 300 °C. The sampling cone voltage and source offset were set to 20 V and the source temperature was 120 °C. The measurements in negative ion mode were performed with capillary voltage set to 3.00 kV. The desolvation gas flow was 700 L/h and temperature 300 °C. The sampling cone voltage and source offset were set to 20 V, and the source temperature was 120 °C. Samples were dissolved in methanol or methanol-$d_4$, degassed and injected directly into the electrospray ion source. Methanol was used as a mobile phase with the flow rate 100 µl/min. The instrument worked with external calibration of sodium formate in the mass range of m/z = 50-1200. The Leucine-Enkephaline solution was used as the Lock-Spray reference material. The lock spray spectrum of the leucine-enkephalin was generated by the lock spray source and correction was done for every spectrum. The exact mass measurements for all peaks were performed within 3 mDa mass error.

### 4.2. Computational Methods

The ground-state equilibrium geometries of the compounds considered herein were determined with the second-order Møller-Plesset (MP2) ab initio method.[27] Vertical electronic excitation energies of singlet and triplet excited states were calculated with the algebraic-diagrammatic construction of second order (ADC(2)) method.[28,29] ADC(2) is a computationally efficient single-reference propagator method which yields similar results as the simplified second-order coupled-cluster (CC2) method. The accuracy of CC2 and ADC(2) for excitation energies of organic molecules was extensively benchmarked in comparison with more accurate methods, such as CC3 and CCSDT.[30,31] A mean absolute error of ≈ 0. 22 eV for low-lying singlet states and ≈ 0.12 eV for low-lying triplet states has been estimated. This accuracy is sufficient for the purposes of the present computational study. Additionally, some vertical excitation energies were computed with the time-dependent density functional theory (TDDFT). The TDDFT calculations were performed with widely used exchange-correlation functional, the three parameter Becke, Lee, Yang, Parr functional (B3LYP)[32,33] and with D3 dispersion correction of Grimme.[34]

The reaction path for the proton-coupled electron-transfer (PCET) process from the water or methanol molecule to HATN was constructed as a so-called relaxed scan, that is, for a fixed value of the driving coordinate (the OH bond length of $H_2O$ or $CH_3OH$ molecules with hydrogen involved in the hydrogen bond with nitrogen atom of HATN) all other internal coordinates of the complex were relaxed during geometry optimization in the respective electronic state.

All calculations were performed with the TURBOMOLE program package (V.7.3)[35] making use of the resolution-of-the-identity (RI) approximation. The correlation-consistent polarized valence-split double-ζ basis set (cc-pVDZ)[36] was employed in all calculations with the exception of the reaction path optimizations, where for economic reasons a more compact basis set of TURBOMOLE (def-SV(P)) was used.

## 5 Acknowledgements

We acknowledge grants from Polish National Science Centre: project QuantERA no. 2017/25/Z/ST2/03038 – spectroscopic measurements, project SONATA no. 2019/35/D/ST5/00594 - NMR and electrochemical measurements, and project OPUS no. 2020/39/B/ST4/01723 – theoretical results).

We are also very grateful to Dr. Dorota Staszek and M.Sc. Ewa Gruba of Institute of Organic Chemistry, Polish Academy of Sciences, for performing the mass spectroscopy measurements under oxygen-free conditions.

**Keywords:** Photochemical hydrogen abstraction • hydrogen attachment • hydrogen attachment • Proton Coupled Electron Transfer • alcohol oxidation • water oxidation • EPR radicals detection• mass spectroscopy spectra